\documentclass[structabstract]{aa} 
\usepackage{graphicx,natbib}
\usepackage{txfonts}
\begin{document}
   \title
   {A coincidence between a hydrocarbon plasma 
   absorption spectrum and the $\lambda$5450 DIB}

   \subtitle{}

   \author{
   H. Linnartz, \inst{1}
   N. Wehres, \inst{1,2}
   H. Van Winckel, \inst{3}
   G.A.H. Walker, \inst{4}
   D.A. Bohlender, \inst {5}
   A.G.G.M Tielens, \inst{6}
   T. Motylewski, \inst{7}
   \and 
   J.P. Maier \inst{7}
   }

   \institute{
   Raymond and Beverly Sackler Laboratory for Astrophysics,    
   Leiden Observatory, Leiden University, P.O. Box 9513, NL 2300 RA  
   Leiden, the Netherlands
   \and
   Kapteyn Astronomical Institute, University of Groningen, P.O. Box 800, 9700 AV Groningen, The Netherlands
   \and
   Instituut voor Sterrenkunde, K.U. Leuven, Celestijnenlaan 200B, B  
   3000 Leuven, Belgium
   \and
   Physics and Astronomy Department, University of British Columbia,
   Vancouver, BC V6T 1Z4, Canada 
   \and 
   	National Research Council of Canada, Herzberg
	Institute of Astrophysics, 5071 W. Saanich Road, Victoria, BC V9E 2E7,
	Canada
   \and
   Leiden Observatory, Leiden University, P.O. Box 9513, NL 2300 
   RA Leiden, the Netherlands
   \and
   Department of Chemistry, University of Basel, Klingelbergstrasse 80, 
   CH 4056 Basel, Switzerland
   }

   \date{Received January 08, 2010; accepted xxxx xx, 2010}

 
   \abstract
   {}
   {
   The aim of this work is to link the broad $\lambda$5450 diffuse    
   interstellar band (DIB) to a laboratory spectrum recorded 
   through expanding acetylene plasma. 
   }
   {
   Cavity ring-down direct absorption spectra and astronomical
   observations of HD183143 with the HERMES spectrograph on the
   Mercator Telescope in La Palma and the McKellar spectrograph 
   on the DAO 1.2 m Telescope are compared.
   }
   {
   In the 543-547 nm region a broad band is measured with a band
   maximum at 545 nm and FWHM of 1.03(0.1) nm coinciding with a 
   well-known diffuse interstellar band at $\lambda$5450 with 
   FWHM of 0.953 nm. 
   }
   {
   A coincidence is found between the laboratory and the two
   independent observational studies obtained at higher spectral
   resolution. This result is important, as a match between 
   a laboratory spectrum and a -- potentially lifetime broadened -- 
   DIB is found. A series of additional experiments has been 
   performed in order to unambiguously identify the laboratory 
   carrier of this band. 
   This has not been possible. The laboratory results, however, \ 
   restrict the carrier to a molecular transient, consisting of 
   carbon and hydrogen.}


\titlerunning{Coincidence between a hydrocarbon plasma and the 5450 DIB}
\authorrunning{Linnartz et al.}

   \maketitle

\section{Introduction}

Diffuse interstellar bands are absorption features observed in 
starlight crossing diffuse interstellar clouds. Since their discovery
in the beginning of the 20th century, scientists have been puzzled 
by the origin of these bands that appear both as relatively narrow 
and rather broad bands covering the UV/VIS and NIR 
\citep{Tielens:1995}. In the last decennia the idea has been
established that it is unlikely that all these bands are due to one 
or a very few carriers and with the progress of optical laboratory
techniques, several families of potential carriers have been
investigated. 
It was shown that the electronic transitions of a series 
of PAH-cations do not match the listed DIBs 
\citep{Salama:1996,Salama:1999,Brechignac:1999,Ruiterkamp:2002}.
Similarly, systematic laboratory studies of electronic spectra 
of carbon chain radicals have not resulted in positive 
identifications either 
\citep{Motylewski:2000, Ball:2000a,Jochnowitz:2008}, 
even though it is known from combined radio-astronomical and 
Fourier Transform Microwave (FTMW) studies that many of such 
species are present in dense clouds \citep{Thaddeus:2001}. Only 
C$_3$ has been recorded unambiguously in diffuse interstellar clouds
\citep{Maier:2001}. 
Other studies, focusing on multi-photon excitation in molecular
hydrogen \citep{Sorokin:1998}, or spectra of fullerenes and 
nano-tubes \citep{Kroto:1992,Foing:1994} have been unsuccessful 
as well. 
In the past years, several coincidences between laboratory and
astronomical DIB studies have been reported in the literature. 
These have all turned out to be accidental, and from a statistical 
point of view, the chance of an overlap is also quite substantial, 
 DIBs cover a major part of the wavelength region between 
roughly 350 and 1000 nm. However, there are a number 
of conditions that have to be fulfilled before a coincidence 
of a laboratory and an astronomical DIB spectrum may be interpreted 
as a real match. These conditions have become more and more strict 
with the recent improvement in achievable spectral resolution, both 
in laboratory and astronomical studies.  

   \begin{figure}
   \centering
   \includegraphics[width=9cm]{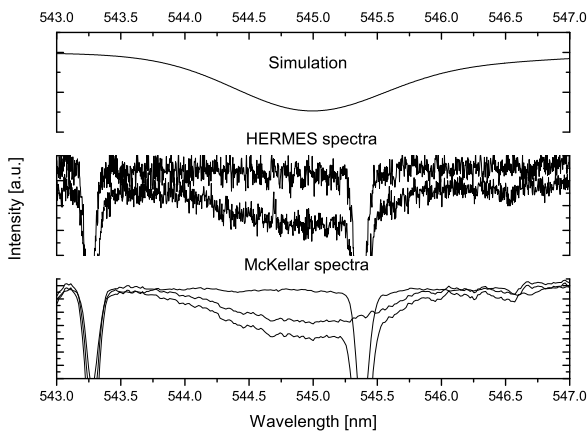}
   \caption{The $\lambda$5450 DIB. The top spectrum is a simulated 
   spectrum as available from DIB catalogues. The middle and bottom
   figure show observational spectra of the HERMES and McKellar
   spectrograph, respectively.
   The HERMES spectra show the $\lambda$5450 DIB recorded
   toward HD 183143 and toward a reference star (HD 164353). The
   McKellar spectra show a reference spectrum toward Rigel (top),
   the DIB spectrum also toward HD 183143 (bottom) and the corresponding
   spectrum (middle) in which the SII stellar line has been deblended. 
   }
   \label{observational spectra}
   \end{figure}

The two most important DIB matching criteria to link laboratory and 
astronomical data are:\\

\begin{enumerate}
\item The gas phase laboratory and observational values of both peak 
position and bandwidth of the origin band transition should be
identical, unless it can be argued that a spectral shift or band
profile change may be due to an isotope or temperature effect. 
An example of the latter is given by spectroscopic measurements
on benzene plasma yielding an absorption feature coinciding with the
strongest DIB at 442.9 nm \citep{Ball:2000b,Araki:2004}. 
The laboratory FWHM turned out to be narrower than in the astronomical
spectrum. It was argued that the spectrum of a non-polar molecule 
cooled in a molecular expansion may be considerably colder
than in space where only radiative cooling applies. A similar
discussion has been given by \citep{Motylewski:2000} who showed that
unresolved rotational profiles may change substantially for different
temperatures, as has also been calculated and discussed by 
\cite{Cossart:1990}.  
\\
\item Once the origin band overlaps with a DIB feature, gas phase 
transitions to vibrationally excited levels in the electronically
excited state of the same carrier molecule should match as well and 
the resulting band profiles should behave in a similar way (i.e. with
comparable equivalent width ratios) \citep{Motylewski:2000}.
A good example for this is the electronic spectrum of C$_7$$^-$ that
has been regarded for several years as a potential carrier as
subsequent electronic bands fulfilled both conditions   
\citep{Tulej:1998,Kirkwood:1998}. Detailed follow-up studies showed
that the series of (near) matches was coincidental \citep{McCall:2001}. \\
\end{enumerate}
In the end, and despite much progress both from the observational and
laboratory side, all efforts to assign DIBs have resulted in a rather
static situation -- triggering more and more exotic explanations for
DIB carriers -- and the origin of the DIBs is still as mysterious as it 
was nearly 100 years ago. 

In this letter we report a match of a laboratory spectrum with a 
diffuse interstellar band that is special as the first condition 
is fulfilled for a rather broad and potentially lifetime broadened 
DIB, i.e. the laboratory and astronomical spectrum should be fully
identical, independent of temperature restrictions. New astronomical
observations obtained with the Mercator telescope, using
the HERMES spectrograph and the Dominion Astrophysical Observatory
(DOA) 1.2 meter telescope, using the McKellar spectrograph are
presented in order to characterize the band profile of the 
$\lambda$5450 DIB with the best possible resolution.
Even though we have not been able to unambiguously identify 
the laboratory carrier, that most likely is a smaller hydro-carbon
bearing molecular transient, we think that this overlap is important 
to report, as it provides a new piece of the puzzle. 

\section{Laboratory Experiments}

The experimental set-up has been described 
\citep{Linnartz:1998, Motylewski:1999} and has been extensively used 
to study a large number of carbon chain radicals of astrophysical
interest \citep{Jochnowitz:2008}. The monochromatic output $\sim$ 
0.1\ cm$^{-1}$ at 540 nm ($\sim$ 18,500\ cm$^{-1}$) of a pulsed dye 
laser based cavity ring-down setup is focused into an optical cavity
consisting of two highly reflective mirrors (R $>$ 0.9999). A 
special pulsed high pressure slit-nozzle system capable of producing
intense 300 $\mu$s long plasma pulses by discharging (- 1 kV, 100 mA)
an expanding gas mixture of 1 $\%$ acetylene (C$_2$H$_2$) in He
is mounted inside the cavity with its slit parallel to the optical 
axis of the cavity. In the expansion a large variety of new species 
is formed and as the technique is not mass selective, special care 
has to be taken when assigning bands to specific carriers. Mass
selective matrix isolation spectra offer a good starting point for 
an assignment \citep{Jochnowitz:2008}. In the case of rotationally
resolved spectra unambiguous identifications are generally possible,
either by combination differences of accurate spectral fits, or by
isotopic studies using C$_2$D$_2$ instead of C$_2$H$_2$ (or a mixture
of C$_2$H$_2$/C$_2$D$_2$). The source runs at 30 Hz and special care is
taken that the pressure inside the cavity remains constant during jet
operation to reduce baseline fluctuations. Rotational temperatures 
are typically of the order of T$_{rot}$ $\sim$ 10\--20 K. This low
temperature results in a spectral simplification and simultaneously
increases the detection sensitivity because of an improved 
state-density. In addition, the source offers a Doppler free
environment with a relatively long effective absorption path length.
The laser beam intersects the 3 cm long  planar expansion about 
5\--10 mm downstream using a sophisticated trigger scheme. Subsequent
ring-down events (typically 20-30 $\mu$s for a 52 cm long cavity) 
are recorded as function of the laser frequency by a photo-diode 
and transferred to an averaged ring-down time by fitting 45 
subsequent ring-down events. This value as function of the 
laser wavelength provides a sensitive way to record optical spectra. 
An absolute frequency calibration is obtained by recording an I$_2$
reference spectrum simultaneously. 

\section{Astronomical Observations}
The laboratory data are compared to observations from two different
astronomical facilities.
\subsection{HERMES @ Mercator Telescope}
The HERMES observations were carried out in service mode using the 
Mercator telescope at Roque de los Muchachos Observatory on La Palma.
The 1.2 m telescope is operated by the Katholieke Universiteit in
Leuven, Belgium, in collaboration with the Observatory in Geneva, 
Switzerland.   
The spectra were obtained in June 2009 with HERMES (High Efficiency 
and Resolution Mercator Echelle Spectrograph) \citep{Raskin:2008},
which is a fibre-fed-cross-dispersed spectrograph.  
The spectrograph has a fixed spectral format and samples the spectrum
between 377 and 990 nm in 55 spectral orders on a 4.6\ k x 2\ k CCD.
The spectral resolution is slightly variable over the field, 
but is 85,000 on average. We obtained 3 spectra of 1,200\ s of 
HD 183143 (B7Ia, m(v)=6.92, B-V=+1.001), the DIB spectral standard 
with a reddening E(B-V) close to 1.0. The reference star HD 164353 
(B5Ib, m(v)=3.97, B-V = $-$0.002) was sampled in 3 exposures of 1 min.
The spectral reduction was performed using the specifically coded
HERMES pipeline and contains all standard steps in spectral reduction.
The wavelength calibration is based on spectra of ThAr and Ne lamps.
As we are mostly interested in the broad absorption feature that 
is centred around 545 nm, we focus further on this spectral region 
of HD 183143.  The spectra are shown in Figure 1 (middle rows) and
compared to the $\lambda$5450 DIB profile as available from a 
series of digital DIB catalogues 
\citep{Herbig:1975, Jenniskens:1994, Tuairisg:2000, Galazutdinov:2000} 
in the upper row.

\subsection{McKellar @ DAO Telescope}
Fifty-five half-hour spectra were taken with the McKellar Spectrograph
and SITe-4 CCD at the DAO 1.2 m telescope, operated by the National 
Research Council of Canada, over 6 nights between 16 and 23 July 2006
(UT) at a dispersion of 10.1 $\AA$/mm giving 0.151 $\AA$/pixel
for a resolution $\sim$ 0.3 $\AA$.  The data were processed in a 
standard fashion using IRAF
\footnote{IRAF is distributed by the National
Optical Astronomy Observatory, which is operated by  the Association of
Universities for Research in Astronomy (AURA) under cooperative agreement
with the National Science Foundation.}. 
The aggregate spectrum had a 
signal to noise of about 1200/pixel before correction of telluric lines. 
Removal of the quite weak telluric features was performed conventionally 
with spectra (S/N $\sim$1600) of the A0 V star zeta Aql  (HD
177724) as the template.

Rigel, an unreddened comparison star with a B8 Ia spectral type very
similar to the B7 Ia of HD 183143, was also observed in order to identify
photospheric lines which contaminate the interstellar features observed in
the latter star.  The sharp line at approximately 5454\AA\ arises from S
II and was removed from the spectrum of HD 183143 by simply fitting a
Voigt profile to the line and subtracting this from the original spectrum.
 The final ``deblended'' spectrum is plotted as a comparison in Fig. 1
(lower panel, middle spectrum).

\section{Results}
In Fig. 2 several spectra in the 543-547 nm region are compared. 
The top spectrum is the digital DIB spectrum of the $\lambda$5450 
DIB \citep{Herbig:1975, Jenniskens:1994, Tuairisg:2000, Galazutdinov:2000}. 
The spectrum in the middle is a zoom in on the 
deblended McKellar spectrum as shown in Fig. 1. The bottom spectrum 
is the laboratory spectrum recorded in direct absorption through 
an expanding 1$\%$ C$_2$H$_2$/He plasma. The similarity between 
the three spectra is striking. 

This wavelength region was initially scanned to search for the 
$^1$$\Pi$$_u$\ --\ X$^1$$\Sigma$$_g$$^+$ electronic origin band
spectrum of the linear carbon chain radical C$_7$ (following the 
C$_7^-$ DIB discussion) that was located in matrix isolation
experiments around 542.3 nm. The laboratory spectrum, shown in 
Fig. 2, consists of many narrow lines that are due to small 
acetylene fragments (typically C$_2$ and CH) that get weaker 
when the distance from the nozzle orifice to the optical axis is increased, 
but there is clearly a broad feature lying underneath. 
As this band shifts by 1.5 nm to the red upon C$_2$D$_2$ precursor
substitution, it initially was neglected, as for C$_7$ both 
C$_2$H$_2$ and C$_2$D$_2$ should result in an identical spectrum. 
The shift is illustrated in Fig. 3. In addition, the deuterated spectrum 
appears to be somewhat stronger. Despite this
negative result for C$_7$ the profile hiding under the narrow 
lines in the C$_2$H$_2$ precursor experiment perfectly matches the 
$\lambda$5450 DIB as available from the DIB databases, reason 
why additional observations were performed.
 
   \begin{figure}
   \centering
   \includegraphics[width=9cm]{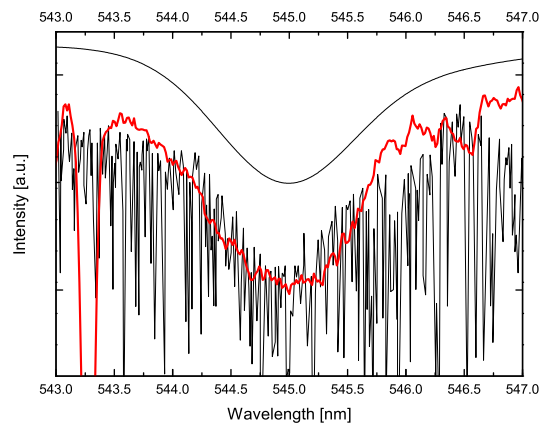}
   \caption{The top spectrum shows the digital $\lambda$5450 DIB, the
   middle spectrum shows the deblended McKellar data and
   the bottom spectrum shows the laboratory cavity ring-down absorption
   spectrum through a supersonically expanding acetylene plasma.}
   \label{}
   \end{figure}

   \begin{figure}
   \centering
   \includegraphics[width=9cm]{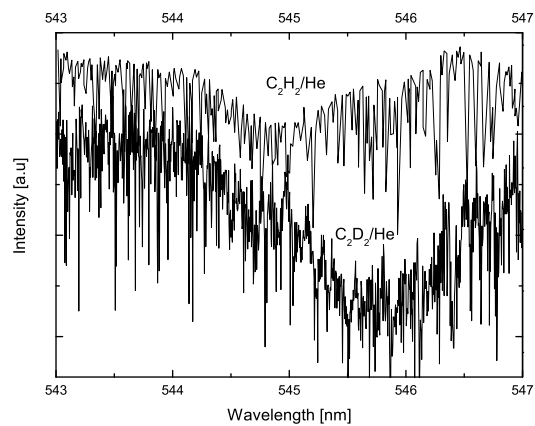}
   \caption{Comparison between laboratory experiments sampling 
   expanding plasma using regular acetylene (top) and deuterated 
   acetylene (bottom) as a precursor gas.}
   \label{observational spectrum and simulation}
   \end{figure}
 
\section{Discussion}

There is little discussion possible about the coincidence between the
recorded laboratory spectrum and the $\lambda$5450 DIB. Both bands
have a central peak position of 545 nm and a FWHM of 1.03 (0.1) nm 
(laboratory spectrum) and 0.953 nm (observational spectrum) 
\citep{Tuairisg:2000}. The uncertainty in the first value is due to 
the overlap of the many individual transitions that prohibits a clear
view on the broad feature. The question is more whether this actually
represents a DIB match and for this complementary information is
needed. Additional laboratory work has been performed, where it 
should be noted that scans as shown in Figs. 2 and 3 typically last 
45 minutes to an hour, in order to achieve the required sensitivity 
and to cover a frequency domain large enough to discriminate band
profile and base line, i.e. fast optimizations are not possible.
The laboratory band does not show any structure that can be related 
to unresolved P, Q and R-branches. With 1.03 (0.1) nm the band is 
also much broader than the unresolved rotational profile of a larger
carbon chain radical. For comparison, at 15 K, the band profile of 
the linear C$_6$H radical (at 525 nm) is about five times narrower
\citep{Linnartz:1999}. It should also be noted that such a broad
feature actually represents a large absorption compared to many of 
the sharper DIBs. Changing the experimental settings to vary 
the final temperature in the expansion by measuring close 
($\sim$ 50\ K\ $^{\prime}$warm$^{\prime}$) and far 
($\sim$ 10\ K\ $^{\prime}$cold$^{\prime}$) downstream, does not
substantially change the FWHM of the spectral contour. 
As the narrow overlapping transitions have FWHMs close to the 
laser bandwidth, experimental broadening artifacts such as residual
Doppler broadening in the expansion or amplified spontaneous emission,
can be excluded. It is clear that the band profile is due to a
temperature independent and carrier specific broadening effect,
presumably life time broadening. The observed bandwidth of 1.0 nm 
($\sim$ 35\ cm$^{-1}$ around 545 nm) corresponds to a lifetime 
of roughly 0.15 ps.
The bandwidth profile does not allow concluding on the nature of 
the laboratory carrier. The carrier must be a transient species 
(a molecular radical, a cation or anion, a weakly bound 
radical complex, possibly charged, or a vibrationally or 
electronically excited species) as no comparable spectra are 
recorded without plasma (i.e. with a regular C$_2$H$_2$/He 
expansion). The use of a C$_2$D$_2$/He expansion results in 
a red--shifted spectrum (Fig. 3) and from this it can be concluded 
that the laboratory carrier must contain both carbon and hydrogen. 
In order to check whether there are equivalent H-atoms in this 
carrier a C$_2$H$_2$/C$_2$D$_2$ 1:1 mixture in He has been used 
as an expansion gas, but this only results in a very broad 
absorption feature covering the whole region between results 
obtained from pure C$_2$H$_2$ and pure C$_2$D$_2$ expansions. 
It is not possible, as demonstrated for HC$_6$H$^+$ or HC$_7$H 
\citep{Sinclair:1999, Ball:2000a, Khoroshev:2004}, to conclude 
on the actual number of equivalent H-atoms in the carrier by 
determining the number of bands that shows up.  Also the use
of another precursor (e.g. allene) did not provide conclusive information.
 
Additional experiments have been performed. The 543-545 nm region 
has been scanned using a two-photon REMPI-TOF experiment with the 
aim to determine the mass of the carrier \citep{Pino:2001}. No 
spectrum could be recorded, which may be related to the short 
lifetime of the excited state or with the fact that the carrier 
is an ion. Ions are indeed formed in this planar plasma source 
\citep{Witkowicz:2004}. Both smaller and larger species have 
been observed, with optimum production rates depending, among 
other things, on the backing pressure. The production of larger species is
generally more critical, e.g. higher backing pressures are needed 
but this also may destabilize the plasma which is unfortunate,
particularly during long scan procedures. More complex species 
are generally found further downstream, but in this specific case 
we did not observe large differences as function of the distance 
from the laser beam to the nozzle orifice. This is the typical
behaviour for a smaller constituent in the gas expansion. We have 
tried to study systematically the voltage dependence of the signal; 
for a positive ion an increase in voltage should go along with a 
decrease in signal for distances further downstream, as the jaws 
carry a negative voltage. For anions it is the opposite, but 10 years
of experience with this source have shown that negative ions are rather 
hard to produce. Again, the changes we recorded were small and did 
not allow drawing hard conclusions. Following condition 2 mentioned 
in the introduction, we have also searched in other wavelength regions 
blue shifted by values typical for an excited C--C, C$=$C, C$\equiv$C
or CH stretch in the upper electronic state. Such excited bands 
have not been observed here, but it should be noted that these 
bands can be intrinsically weak. \\
In summary, we are left with a laboratory spectrum that coincides 
both in band maximum and band width with a known DIB band at 545 nm.
Our measurements show that the absorption spectrum of a transient
molecule containing hydrogen and carbon reproduces the astronomical
spectrum. The profile can be explained with life time broadening 
and this is consistent with the observation that the laboratory 
and astronomical spectrum are identical, i.e. without temperature
constraints. In addition, it explains why the large bandwidth of 
this DIB does not vary along different lines of sight. The large
effective absorption also may be indicative for an abundant carrier. 
The exact carrier, as such, remains an open question. 
The present result, however, may be useful to stimulate upcoming DIB work. 

\begin{acknowledgements}
The results presented here bridge a period of 10 years. The cavity 
ring-down measurements were performed in the Institute for Physical 
Chemistry (Department of Chemistry, University of Basel) with 
support of the Swiss National Science Foundation and the analysis
follows recent observations and a collaboration within the framework 
of the FP6 research training network {\it The Molecular Universe}.
Additional financial support of NOVA is gratefully acknowledged.
\end{acknowledgements}

\end{document}